\documentstyle [12pt,epsf]{article}
\begin{document}
\baselineskip=20.0pt
\begin{titlepage}
\vspace{1cm}
\begin{center}
{\bf {\Large Short-time critical dynamics and universality}}\\
\vspace{0.2 cm}
{\bf {\Large on a two-dimensional Triangular 
Lattice\footnote{Work supported in part by the National Natural Science Foundation 
of China, 19975041 and 10074055.
}}}

\vspace{1.0 cm}
{\bf \large H.P. Ying$^{a,b}$, L. Wang$^{a}$, J.B Zhang$^{a}$, 
M. Jiang$^{a}$ and J. Hu$^{a}$}

\vspace{0.2 cm} 
{\small 
$^{a}${\it  Zhejiang Institute of Modern Physics, Zhejiang University,
Hangzhou 310027, P.R. China }\\ }
\vspace{0.2cm}
{\small
$^b${\it FB Physik, Universit\"at-GH Siegen, 57068 Siegen, Germany ~~~}\\
}
\end{center}
 
\vspace{0.6cm}
\begin{abstract}
Critical scaling and universality in short-time dynamics for 
spin models on a two-dimensional triangular lattice are investigated 
by using Monte Carlo simulation. 
Emphasis is placed on the dynamic evolution from fully ordered initial
states to show that universal scaling exists already in 
the short-time regime in form of power-law behavior of the
magnetization and Binder cumulant. The results measured for the
dynamic and static critical exponents, $\theta$, $z$, $\beta$ and 
$\nu$, confirm explicitly that the Potts models on the triangular
lattice and square lattice belong to the same universality class. 
Our critical scaling analysis strongly suggests that the simulation 
for the dynamic relaxation can be used to determine numerically
the universality.
\end{abstract}

\vspace{0.2cm}

\vspace{0.1cm}
\vspace{0.8cm}
 
\end{titlepage}

\section{Introduction}

The study of phase transitions and critical phenomena has been an attractive 
and important topic in statistical physics for a long time \cite{Bind92}.
As is well known there appears critical universality due to the 
infinite spatial and time correlation lengths at the continuous phase 
transition points, and the universal behavior of a critical system is 
characterized by a number of critical exponents such that the models 
in the same universality class have the same values of the critical
exponents. Therefore, an estimation of critical exponents to determine
the universality class for different statistical models is an  
interesting challenge. However such research is difficult to be 
carried out analytically since only few statistical 
models can be exactly solved \cite{Baxter82}. Therefore, numerical 
simulations supply a powerful method to estimate critical exponents 
which are measured by generating equilibrium states and averaging over 
independent configurations in Monte Carlo (MC) studies. For example, 
the dynamical exponent $z$ can be measured from the exponential decay 
of time correlation for finite systems in the long-time regime. 
Unfortunately, MC simulations near the critical point in the equilibrium 
suffer from critical slowing down (CSD) \cite{Swen87}, because the 
auto-correlation length $\tau$ divergs as $\tau \sim L^z$, where $L$ 
denotes the linear size of a lattice and $z \geq 2$.

In the last decade, the exploration of critical phenomena has been 
greatly broadened and much progress has been made in critical dynamics.
An important discovery is that there exist critical scaling 
and universality for systems far from equilibrium \cite{Zheng98}. 
Traditionally, it was believed that universal behavior exists only 
in the equilibrium or in the long-time regime of dynamical 
evolutions. However, recent researches in critical dynamics for 
many statistical models have shown that universal scaling behavior 
also emerges within the macroscopic short-time regime of dynamic 
processes after a microscopic time scale $t_{mic}$ \cite{Janss89,Huse89}. 
The investigation of the {\it short-time 
critical dynamics} (STCD) not only exhibits the existence of universal 
dynamic scaling behavior within the short-time regime, but also supplies
a very efficient method to determine the critical exponents. 

In Ref.\cite{Zheng98}, the recent progress in short-time dynamical 
studies has been reviewed, where several simulation methods and the 
corresponding numerical results have been presented.
Up to now, the values of critical exponents for a variety of statistical 
models, such as the  two-dimensional (2d) Potts and $\phi^4$ models on a square 
lattice \cite{Li95,Okano97,Zheng99}, the 2d quantum XY and the random-bond 
Potts models on the square lattices \cite{Ying98,Ying00}, the 2d FFXY model and 
(2+1)d SU(2) lattice-gauge models \cite{Luo98,Okano98}, 
have been calculated precisely. But only very few attention has been  
concentrated on the 2d triangular lattices \cite{Diehl96,Ritsch95}.

By the STCD method we can estimate not only the dynamic critical 
exponent $z$, but also the static critical exponents $\theta$,
$\beta$ and $\nu$. More important, the results for both the 
dynamic critical exponent $z$ and static critical exponents, $\beta$ 
and $\nu$, are consistent with those obtained by the traditional 
MC simulations performed in equilibrium. Furthermore, similar to
the measurements of critical exponents, the determination of critical 
temperatures is also difficult in equilibrium because of the CSD. 
Now we have an alternative approach, the STCD, by which the critical 
temperature can be also extracted from the power-law scaling behavior in
the critical regime \cite{Okano97}. Therefore, we can perform comprehensive 
investigations on statistical models by the STCD approach which are 
independent of the study in equilibrium.

In this paper we present our numerical study for the $q$-state Potts model 
on a 2d triangular lattice by using the STCD method. Generally, in  
short-time MC simulations, there are two different schemes to calculate 
the critical exponent: one is based on the power-law behavior and the other 
on the finite size scaling collapse for different lattice sizes within
the non-equilibrim relaxation processes. 
Our attention is specially paid to the 
universal short-time evolution from the fully ordered initial state. 
The calculations show evidence that there exists universal scaling already 
in the short-time regime where the power-law behavior of magnetization 
and auto-correlation is observed, as well as the finite size scaling 
collapse behavior. These results are then applied to estimate the critical 
exponents and check the corresponding results for different schemes. 
Our numerical results confirm explicitly the universality 
proposal in the short-time regime \cite{Ritsch97}.

\section{Model and STCD Scaling}

The Hamiltonian for the $q$-state Potts model with ferromagnetic coupling
($J > 0$) defined on a 2d triangular lattice is given by
\begin{eqnarray}
 H = -\frac{J}{2} \sum_{i=1}^{N}\sum_{l_i=1}^6 \delta_{\sigma_i,\sigma_{l_i}},
~~~~\sigma_{i,j}=1,\cdots,q,~~~~~~~~~
\label{ham}
\end{eqnarray}
where $\sum_i$ represents the sum over the lattice sites on a triangular 
lattice $N=L\times L$, and $l_i$ denotes the nearest-neighbors of site $i$, 
whose coordinate number is six on the lattice. $\sigma_i$ is a spin variable 
on every lattice site $i$. It is known that in equilibrium the Potts 
model is exactly solvable, and the critical points are $J_c = 0.5493061\cdots$ 
and $J_c = 0.6309447\cdots$ for $q=2$ and $q=3$ respectively \cite{Baxter82}.
In this work we consider both the $q=2$ and $q=3$ cases and calculate their
critical exponents ($\theta$, $z$, $\beta$, $\nu$)
to investigate the universality explicitly.

As our starting point we consider an $O(n)$ vector model ($n=1$ for the 
Ising model) 
with the dynamics of model A \cite{Hoh77} and let it suddenly quench 
from a very high temperature with small initial magnetization $m_0$ 
to the critical temperature $T_c$. Janssen, Schaub and Schmittmann 
\cite{Janss89} argued that at the critical temperature dynamic scaling 
behavior already emerges within the short-time regime as,
\begin{equation}
~~M^{(k)}(t,\tau,L,m_0)=b^{-k\beta/\nu}M^{(k)}(b^{-z}t,b^{1/\nu}\tau,
b^{-1}L,b^{x_0}m_0),~~~~~~~
\label{general1}
\end{equation}
where $M^{(k)}$ is $k$th moment of the magnetization, 
$\tau=(T-T_c)/T_c$ is the reduced temperature, $\beta$ and $\nu$ are 
the well known static critical exponents and $b$ is a scaling factor, 
while $x_0$, a {\it new independent} exponent, is the scaling dimension 
of the initial magnetization $m_0$. Here a MC sweep over all sites on
the lattice is defined as the unit of MC time $t$. 

For a sufficiently large lattice ($L\rightarrow\infty$) and setting $\tau=0$, 
$b=t^{1/z}$ in the scaling form Eq.(\ref{general1}), the power-law behavior 
of time evolution of the magnetization at the critical temperature can be 
deduced,
\begin{eqnarray}
M(t) \sim m_0 t^\theta,~~~~~~~~~~~~~~ 
\label{mtd}
\end{eqnarray}
where $\theta$ is a new dynamic exponent which characterizes the 
universality in the short-time regime and is related to $x_0$ by 
$\theta = (x_0 - \beta/\nu)/z$. The relation shows that, after the 
microscopic time $t_{mic}$, the magnetization undergoes an initial increase 
at the critical point and we can easily obtain the result of the exponent 
$\theta$ based on this power-law form. 

On the other hand, one can also focus on dynamic processes starting from 
an ordered initial state to estimate exponents $\beta/\nu$ and $z$. 
Although no detailed analytic study has been made about this process, MC 
simulations in a variety of statistical spin models have shown that 
there also exists a similar scaling relation \cite{Okano97,Ito93,Schue95},
\begin{equation}
~~M^{(k)}(t,\tau,L)=b^{-k\beta/\nu}M^{(k)}(b^{-z}t,b^{1/\nu}\tau,
b^{-1}L).~~~~~~~~
\label{general2}
\end{equation}
At the exact critical temperature($\tau=0$) and setting $b=t^{1/z}$,
Eq. (\ref{general2}) leads to a power-law decay for the magnetization,
\begin{eqnarray}
M(t) \sim m_0 t^{-\beta/\nu z}, 
\label{mto}
\end{eqnarray}
where $M(t)$ is defined by
\begin{eqnarray}
M(t)&=&\frac{1}{N}<\sum_{i} \sigma_{i}(t)>, ~~~~ (q=2),~~~~~~~~~~~~
\label{m-def-i}
\end{eqnarray}
and
\begin{eqnarray}
M(t)&=&\frac{3}{2N}<\sum_{i}(\delta_{\sigma_{i}(t),1}-\frac{1}{3})>, ~~~~ 
(q=3).~~~~~~~~
\label{m-def-p}
\end{eqnarray}
Here $N = L\times L$ is also the total number of 
spins defined on the lattices and $<\cdots>$ denotes the average over 
independent initial configurations and/or random number sequences.

However, the independent determination of $1/\nu$ seems slightly more 
complicated than $z$ and $\beta/\nu$. We should differentiate 
$\ln M(t,\tau)$ with respect to $\tau$ at the critical point
\begin{eqnarray}
 \partial_{\tau} \ln M(t,\tau)|_{\tau=0} = t^{1/(\nu z)} 
\partial_{\tau'} \ln M(t',\tau')|_{\tau'=0}. ~~~~~~~~~~~~
\label{pm}
\end{eqnarray}
The exponent $1/\nu$ can be determined from this power-law behavior
if the value of $z$ is known.
In the practice of our numerical calculations, the differential to $\tau$ 
is substituted with a reasonable small difference $\Delta \tau$.
  
Furthermore the dynamic exponent $z$ can be determinated independently. 
In order to do so, we introduce a Binder cumulant
\begin{eqnarray}
U(t,L) = \frac{M^{(2)}(t)}{(M(t))^{2}}-1.~~~~~~~~~~~~
\label{udef}
\end{eqnarray}
Here the second moment of the magnetization $M^{(2)}(t,L)$ is defined as
\begin{eqnarray}
M^{(2)}(t) &=& \frac{1}{N^2}<{(\sum_{i} S_{i}(t))}^2>, ~~~~ (q=2),~~~~~~~~~~~~
\label{cont17}
\end{eqnarray}
and
\begin{eqnarray}
M^{(2)}(t) &=& \frac{9}{4N^2}<(\sum_{i} (\delta_{\sigma_{i}(t),1}-
           \frac{1}{3}))^2>, ~~~~ (q=3),~~~~~~~~~~~~
\label{cont20}
\end{eqnarray}
respectively.
For sufficiently large lattices, finite size scaling analysis
shows a power-law increase of the Binder cumulant, 
\begin{eqnarray}
U(t,L) \sim t^{d/z}.~~~~~~~~~~~~
\label{u}
\end{eqnarray}
Based on this feature the dynamic exponent $z$ can be determined
and then be used to calculate the exponent $1/\nu$ through
Eq.(\ref{pm}). 

As done by many authors \cite{Li95,Ying00,Schue95}, the critical exponents 
can be measured both from disordered initial states and ordered initial states. 
We perform our investigations involving these two different initial 
states for MC measurements to obtain more reliable results.
Particularly we paid much attention to the scaling forms which describe 
the dynamic processes from the ordered initial state instead of the disordered 
initial states to avoid the finite $m_0$ effect on the measured results. 
There is also an advantage to reduce the statistical fluctuations for 
determination of critical exponents and critical temperatures from ordered 
initial dynamic processes. 

Then, there is another scheme to estimate these critical 
exponents, $\beta/\nu$ and $z$, based on the finite size scaling 
collapse \cite{Li95}. The results estimated should be consistent 
with those obtained by the power-law behavior
described above. This scheme gives an independent 
check whether our dynamic MC simulations are reliable or not.
Set $\tau=0$ and $m_0=0$, according to the dynamic scaling behavior
Eq. (\ref{general1}), the simple scaling relation for the second moment
magnetization $M^{(2)}(t,L)$ is easily obtained
\begin{eqnarray}
M^{(2)}(t,L) = b^{-2\beta/\nu} M^{(2)}(b^{-z}t,b^{-1}L).~~~~~~~~~~
\label{M2scaling}
\end{eqnarray}
Using this scaling relation Eq.(\ref{M2scaling}) 
for a pair of curves of $M^{(2)}(t,L)$ and $M^{(2)}(t',L')$ with 
$t'=t/2^z$ and $L'=L/2$, we can estimate the critical exponents 
$\beta/\nu$ and $z$ by searching for the best fit of collapse of the 
curve $M^{(2)}(t',L')$ through a global scaling factor $2^{2\beta/\nu}$. 
Also if we input the value of $z$ got 
from the power-law the Eq. (\ref{u}), the value of $\beta/\nu$ can also 
be estimated by the scaling relation Eq.(\ref{M2scaling}).

\section {Simulation and Results}

It has been demonstrated in previous works \cite{Okano97,Luo98a,Zhang99} 
that the results obtained by the Metropolis and Heat-bath algorithms are 
consistent with each other and the latter is somewhat 
more efficient than the former. 
Therefore, in the present paper, the MC simulation is only performed 
by the Heat-bath algorithm. Samples for average are taken over 150,000 
independent initial configurations on the $L^2$ triangular lattices with 
the periodic boundary conditions using $L=32, 64$ and 128. 
Statistical errors are simply estimated by performing three groups of 
averages selecting different random seeds for the initial configurations. 
Our simulation is performed at or near the critical temperature.

We begin the study by considering the dynamic process starting from random 
initial states with small $m_0$ to determine the critical exponent $\theta$. 
These disordered initial configurations with given values of $m_0$ are 
prepared by the sharp preparation method \cite{Okano97}, and the evolutions 
of the magnetization $M(t)$ are measured as a function of MC time $t$ for 
$m_0=0.04, 0.02$ and $0.01$ on the $N=128^2$ lattice.
The curves of $M(t)$ in a log-log scale are shown in Figs. 1 and 2 for 
the $q=2$ and $q=3$ model, respectively. Obviously, there exist 
very nice power-law increases of $M(t)$ after $t_{mic} \simeq 10$ and all 
the curves are almost parallel to each other. Thus the $\theta$ can be 
estimated from the slopes of the curves in the regime of $t$=[10, 200]. 
In Table 1, the values of $\theta$ as a function of 
initial magnetizations $m_0$ 
are presented for $q=2$ and $q=3$. Then the final results of 
$\theta$, after an extrapolation to $m_0 \rightarrow 0$, are summarized 
in Table 2 both for the models on the triangular and square lattice. 
It is found that the values of $\theta$ on the triangular lattice are the 
same as those on the square lattice.
This calculation shows first evidence of the dynamic universality
in the short-time regime.

Secondly we study the evolution of magnetizations in the initial stage of 
the dynamic relaxation starting from fully ordered initial states.
In Figs.3 and 4, the power-law behavior of the Binder cumulant $U(t)$ for 
$q$=2 and $q$=3 on a lattice with $N = 128^2$ is displayed on a log-log 
scale, and the nice power-law of $U(t)$ is clearly exhibited.
The exponent $z$ can be estimated from the slope of the curves: 
$z=2.145(3)$ for $q$=2 and $z=2.148(4)$ for $q$=3. 
In Table 2, the values of $z$ measured for the two models
are presented. For a comparison, we also list in Table 2 
the $z$ values of the corresponding results on the square lattice. 
Then, we investigate the short-time behavior of $M(t)$ 
starting from the fully ordered initial state. The time evolution of 
$M(t)$ for the two models is displayed in Fig.5 and Fig.6 respectively. 
Here it is shown that the finite size effect is small, 
since there is almost no difference for the curves of the
different lattice sizes. From the slopes of these power-law decay curves 
the values of the index $\beta/\nu z$ is determined. 
Then the values of $\beta/\nu$ are estimated by input of the $z$ values.  
The results are listed in Table 2 for both $q$=2 and $q$=3. 
Our measured results of $\beta/\nu$ are more accurate than those obtained 
in the previous works, compared with the exact values of $\beta/\nu=1/8$ 
for the $q$=2 (Ising) and $\beta/\nu=2/15$ for the $q$=3 Potts model. 
 
Next the MC simulations for $\partial_{\tau} \ln M(t)$ are carried out by 
taking the difference $\Delta \tau$ =0.02 on $L$=128, 64 and 32 lattices. In 
Figs.7 and 8 we plot the evolution of $\partial_{\tau} \ln M(t)$ for
the two models respectively. The nearly complete overlap of these curves on the 
different lattice sizes indicates that the finite size effect can be ignored. 
We also notice that the power-law behavior is not yet seen before 
$t\approx20$, as the effect of microscopic time scale $t_{mic}$. 
Therefore, we measure 
the exponent $1/\nu z$ from the interval [50,500], and the measurements  
yield the results $1/\nu$=1.027(6) for the $q$=2 model and 
$1/\nu$=1.223(5) for $q$=3. These results are included
in Table 2 and compared with those on the square lattice. 

Finally, we study the time evolution of $M^{(2)}(t,L)$ with completely
random initial states ($m_0$=0) on lattices up to $L=128$
to estimate the critical exponent $\beta/\nu$ and $z$ by the finite size 
scaling collapse scheme, Eq.(\ref{M2scaling}), as an alternative way. 
Figs.9 and 10 show the time evolution of the second moment of the magnetization 
$M^{(2)}(t,L)$ for $L=128,64,32,16$ by solid lines. The dots fitted 
to the lines show the data collapse for the rescaled variables $L'=L/2$, 
$t'=t/2^z$ and global scaling factor $2^{2\beta/\nu}$, 
with the standard least square fitting algorithm.  In other words,
we rescaled the $M^{(2)}(t,L)$ lines on a larger lattice to a smaller lattice 
$L'=L/2$ by the scaling factor $2^z$ and the $2^{2\beta/\nu}$ 
for the pair of different lattice sizes.
The average values for $2\beta/\nu$ and $z$ of $q=2$ and $q=3$ are 
also given in the Table 2.
This step gives a further independent check that the STCD method is
efficient and the results of our dynamic MC simulations are reliable.

\begin{table} \begin{center}
{ \caption{\small The measured values of $\theta$ versus the initial
$m_0$ for the Ising model ($q=2$) and Potts model ($q=3$) on a $N=128^2$
lattice.
The last column gives the results of $\theta$ after extrapolation to
$m_0=0$. }}\vskip 0.5cm
\begin{tabular}{c| c| c| c |c } \hline\hline
$m_0 $    &  0.04    &  0.02    & 0.01     & 0.00      \\ \hline
$q=2$     & 0.183(1) & 0.185(1) & 0.189(2) & 0.191(2)  \\ \hline
$q=3$     & 0.101(1) & 0.089(1) & 0.082(1) & 0.076(2)  \\ \hline\hline
\end{tabular}
\end{center} \end{table}

\begin{table} \begin{center}
{\caption{\small Results for the critical exponents $\theta$, $2\beta/\nu$, 
$1/\nu$ and $z$ for $q$=2 and $q$=3 on 2d triangular lattice. For 
comparison, the values on the square lattice are also listed\cite{Zheng98}. }}
\vskip 0.5cm
\begin{tabular}{c| c| c| c |c} \hline\hline
$       $          &$\theta$  & 2$\beta/\nu$ & $1/\nu$ & $z$       \\ \hline
$q$=2(Triangular)   &&&&  \\
power law results    &0.191(2) & 0.250(5)    &1.027(6) &2.145(3)     \\
scaling collapse results &&0.252(2)&&2.153(2)  \\
(Square, Ref.\cite{Zheng98}) &0.191(1)  &0.240(15)&1.03(2) & 2.155(3)  \\ \hline
exact  &~                   & 1/4&     1  &                  \\  \hline
$q$=3(Triangular)   &&&&  \\
power law results   &0.076(2) & 0.256(6)    &1.223(8) &2.148(4)       \\
scaling collapse results &&0.266(2)&&2.191(1)  \\
(Square, Ref.\cite{Zheng98}) &0.075(3)  &0.269(7)& 1.24(3) &2.196(8)     \\ \hline
exact  &~                   & 4/15&     1.2    &                  \\  \hline
\hline
\end{tabular}
\end{center} \end{table}

\section{Summary and Conclusion}

We have systematically investigated the short-time critical dynamics of
the $q=2$ (Ising) and $q=3$ Potts models on the 2d triangular 
lattices by a large-scale dynamic MC simulation. Firstly, by observing 
the power-law increase of the magnetization $M(t)$ from random 
initial states, the new dynamic exponent $\theta$ has been calculated and 
its values are the same as those for the corresponding models on 
the square lattice. Then we paid our attention particularly to
the time evolutions of the Binder cumulant $U(t)$, magnetization $M(t)$ and 
the derivative $\partial_{\tau} \ln M(t)$ from the fully
ordered initial state to determine the dynamic exponent $z$ and the
static critical exponents $\beta$ and $\nu$. Finally we checked our results 
independently by using the finite-size scaling collapse scheme to the time 
evolutions of $M^{(2)}(t,L)$ from the completely random initial states.     
The advantage of the STCD method, by our experience,
is that the CSD is eliminated because the measurements 
are carried out at the beginning of the evolution process, rather than 
in the equilibrium where all the correlation lengths are divergent. 

In conclusion, the dynamic relaxation in the short-time regime for the 
$q=2$ and $q=3$ Potts models is studied on the 2d triangular lattice,
and it is confirmed that, by comparing our estimated results 
$\theta$, $\beta$ and $\nu$ to those on the 2d square lattice,
they belong to the same universality class 
for the corresponding models on different lattices respectively.
Our critical scaling analysis strongly suggests that the simulation for 
the dynamic relaxations can be used to confirm numerically the universality.
Further application of STCD is interesting to investigate the spin 
systems with quenched disorder \cite{Ying00} and quantum
phase transitions for their universal behavior.

\vspace{0.4 cm}
\noindent {\it Acknowledgments}\\
H.P.Y. would like to thank the Heinrich-Hertz-Stiftung for financial support.
The simulation was performed on the ORIGIN-2K Computer in the Center 
for Simulation and Scientific Computing at the College of Science, 
Zhejiang University.

\newpage

\newpage
\begin{figure}[th]
\setlength\epsfxsize{120mm}
\caption{Plot of $M(t)$ versus $t$ on a log-log scale
for the Ising model on a $N=128^2$ lattice. The lines present the 
different initial magnetizations $m_0$=0.04,0.02,0.01 (from top to bottom).}
\epsfbox{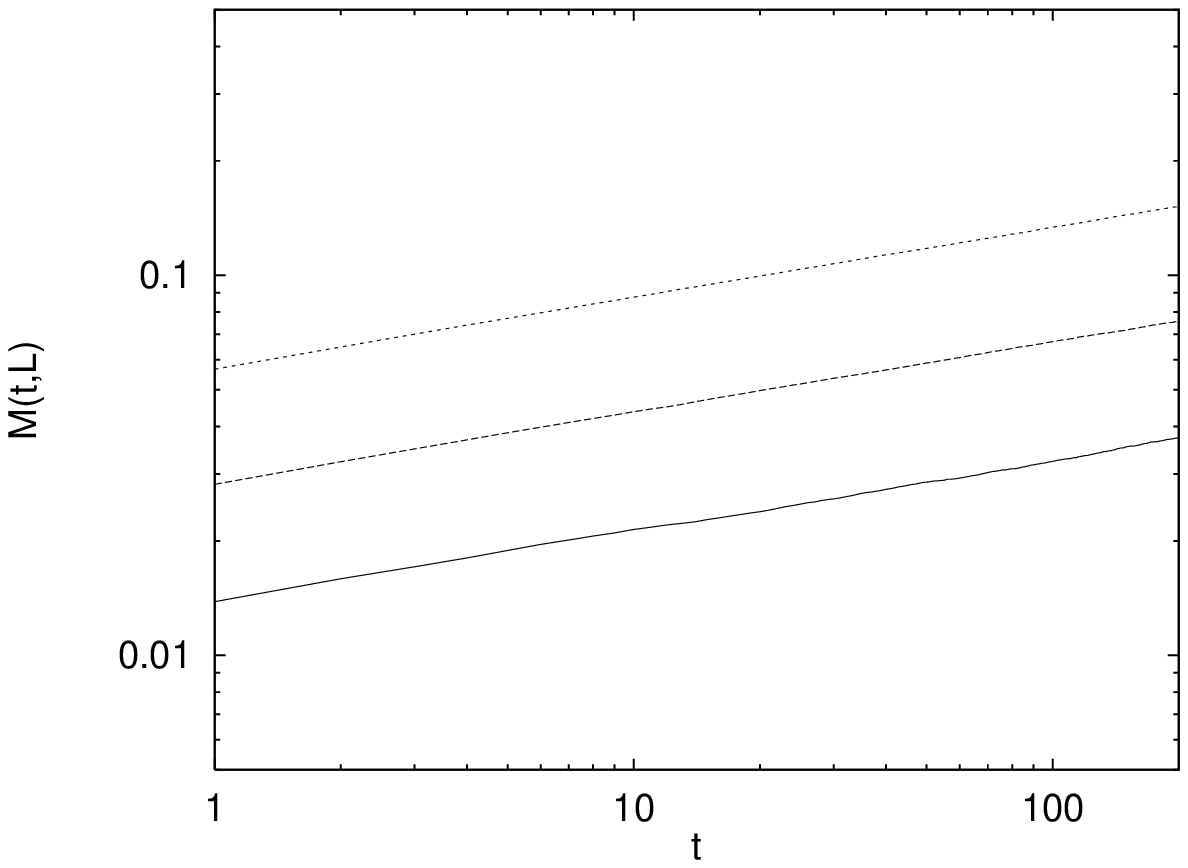}
\vskip 0.4cm
\setlength\epsfxsize{120mm}
\caption{The same as Fig.1, but for the $q$=3 Potts model.~~~~~~~~~~~~~~~~~~~}
\epsfbox{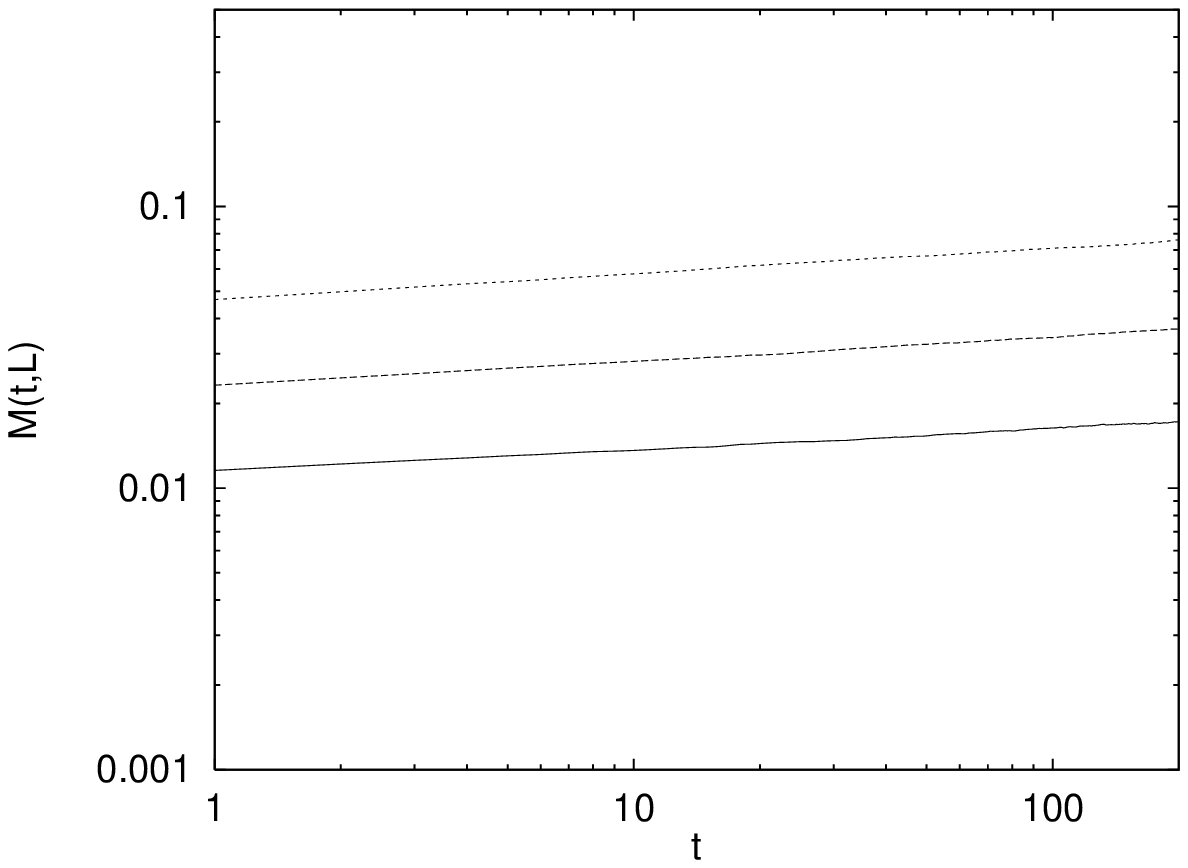}
\end{figure}

\begin{figure}[th]
\setlength\epsfxsize{120mm}
\caption{Time evolution of Binder cumulant $U(t)$ on a log-log scale
for the Ising model on a $N=128^2$ lattice.} 
\epsfbox{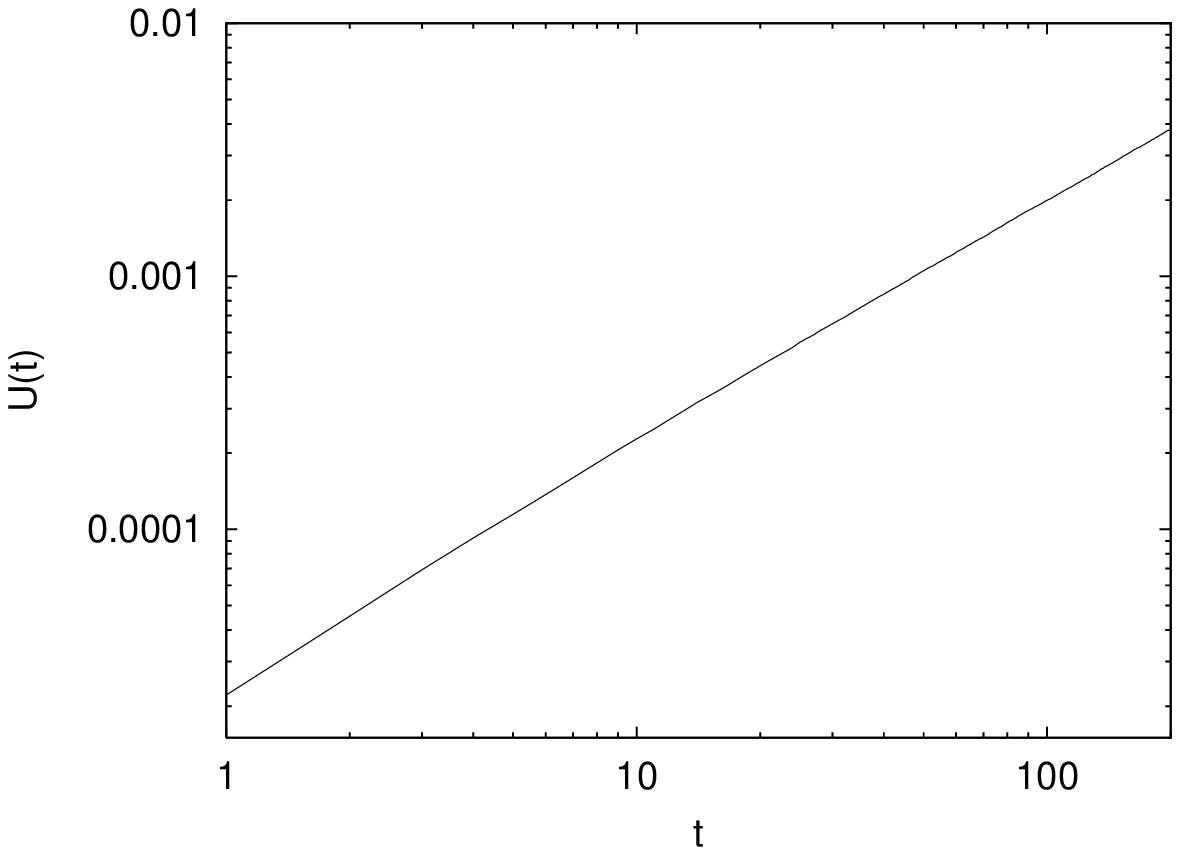}
\vskip 0.4cm
\setlength\epsfxsize{120mm}
\caption{The same as Fig.3, but for the $q$=3 Potts model.~~~~~~~~~~~~~~~~~~~}
\epsfbox{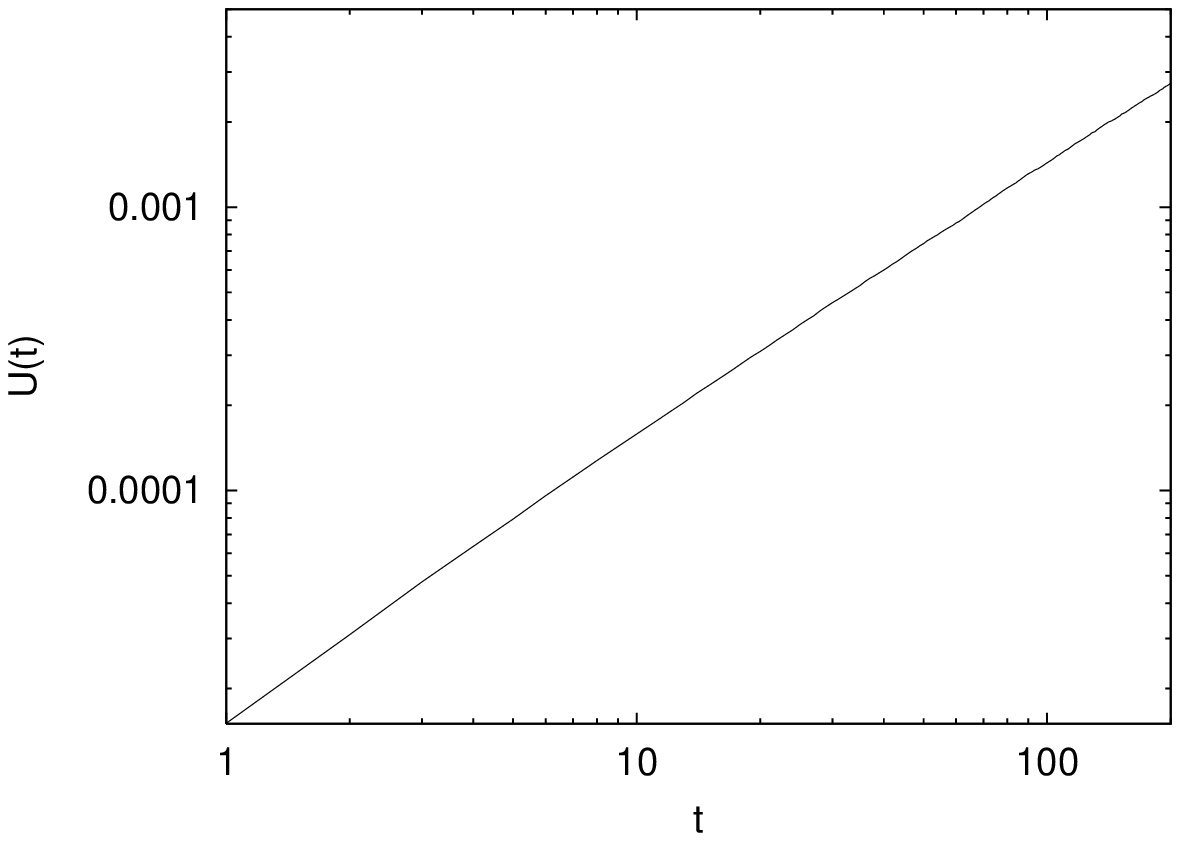}
\end{figure}

\newpage
\begin{figure}[th]
\setlength\epsfxsize{120mm}
\caption{Power-law decay of $M(t)$ versus $t$ on a log-log scale
for the Ising model with $m_0=1$. The lines for different lattices of 
$L$=128,64,32 almost overlap. }
\epsfbox{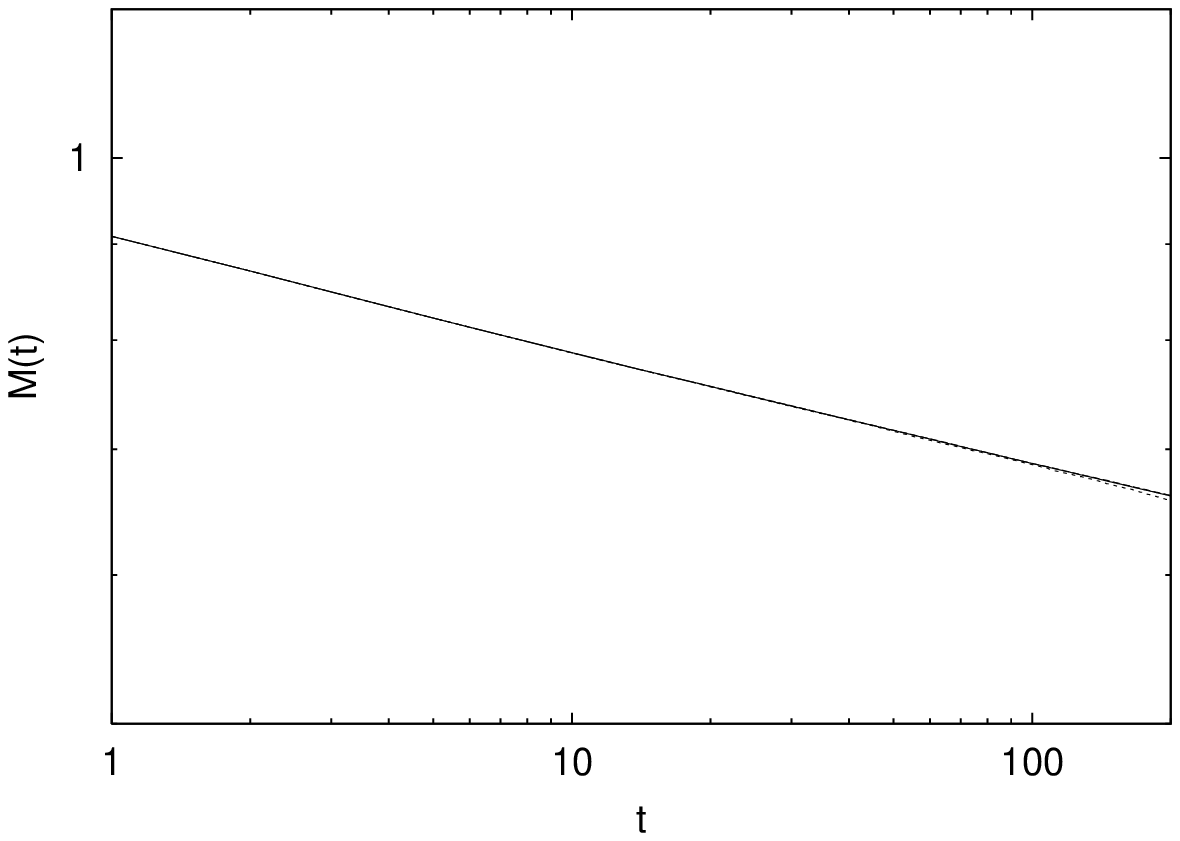}
\vskip 0.4cm
\setlength\epsfxsize{120mm}
\caption{The same as Fig.5, but for the $q$=3 Potts model.~~~~~~~~~~~~~~~~~~~}
\epsfbox{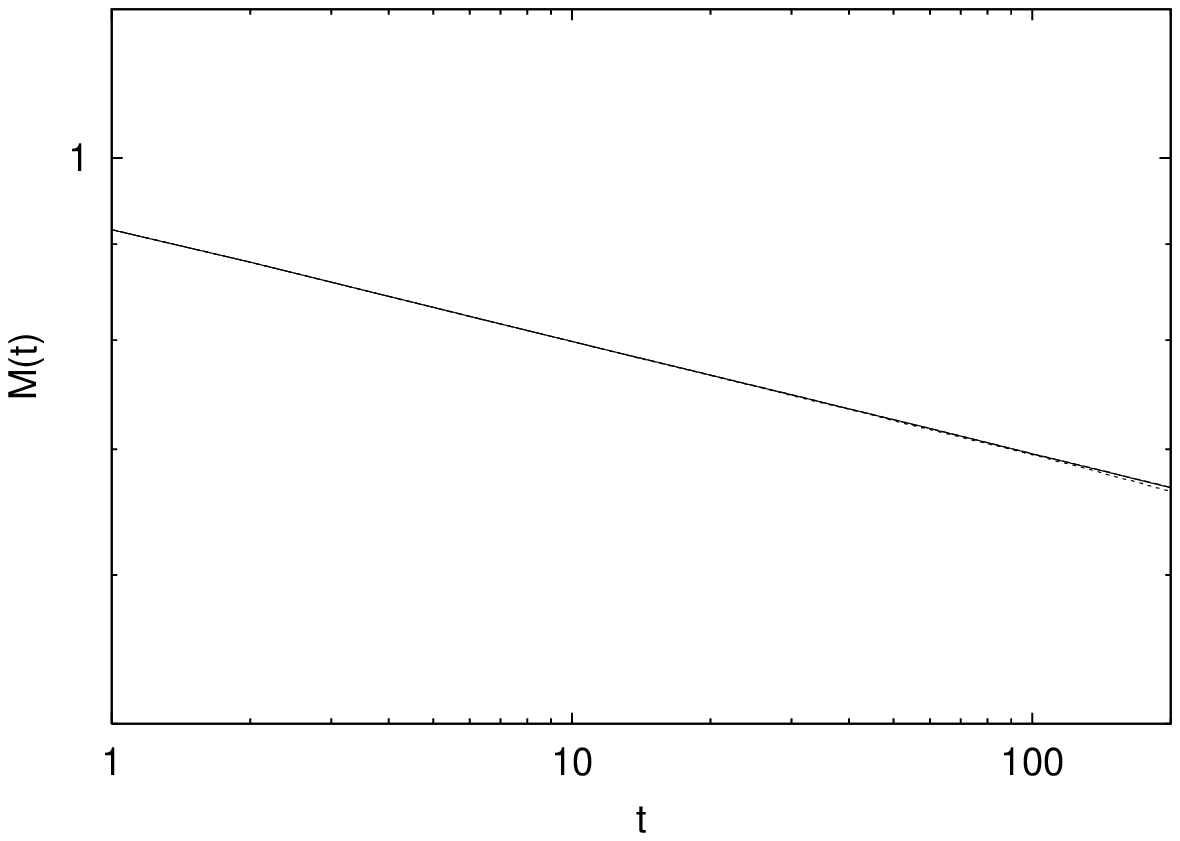}
\end{figure}

\newpage
\begin{figure}[th]
\setlength\epsfxsize{120mm}
\caption{Curves of $\partial_{\tau} \ln M(t)$ on a 
log-log scale for the Ising model with $m_0 =1$ for different 
lattices of $L$=128,64,32.
It is obvious that the finite size effect seems small, and the power-law 
behavior exists after $t_{mic} \simeq 20$. }
\epsfbox{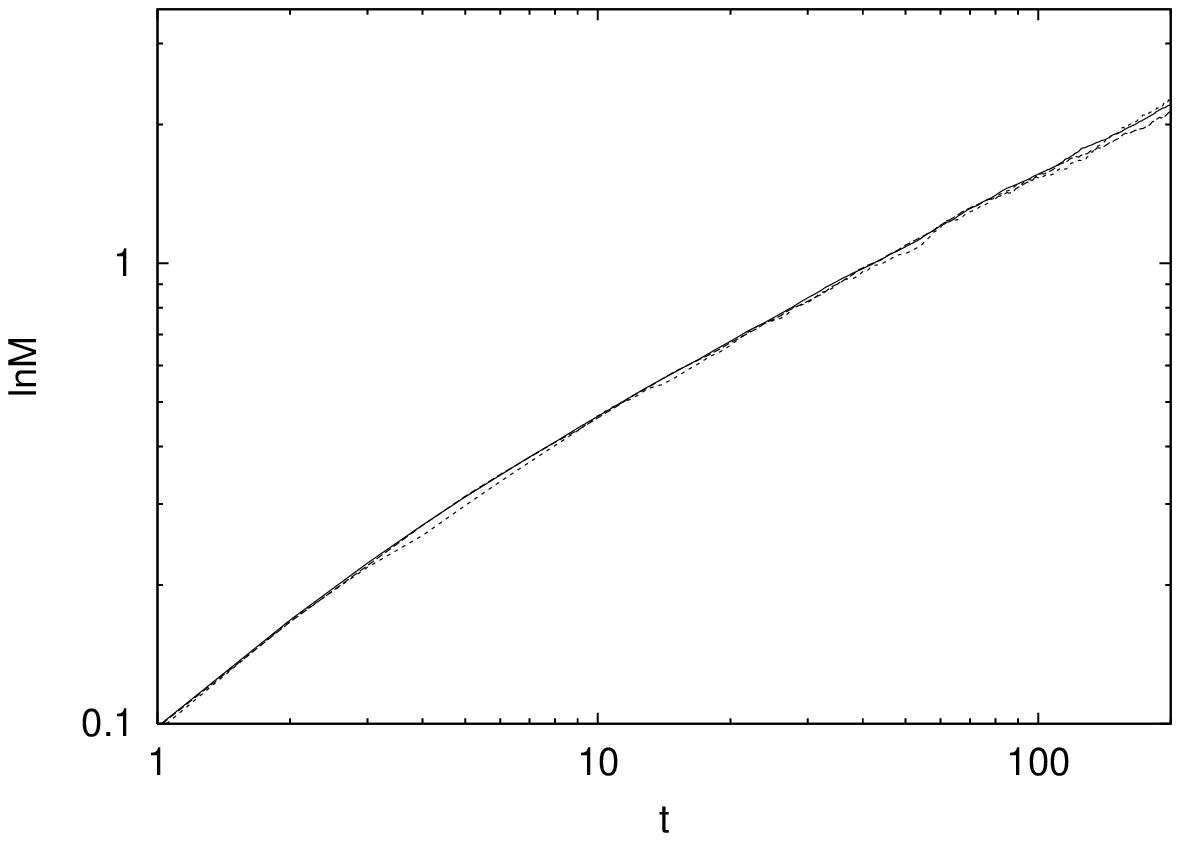}
\vskip 0.4cm
\setlength\epsfxsize{120mm}
\caption{The same as Fig.7, but for the $q$=3 Potts model.~~~~~~~~~~~~~~~~~~~}
\epsfbox{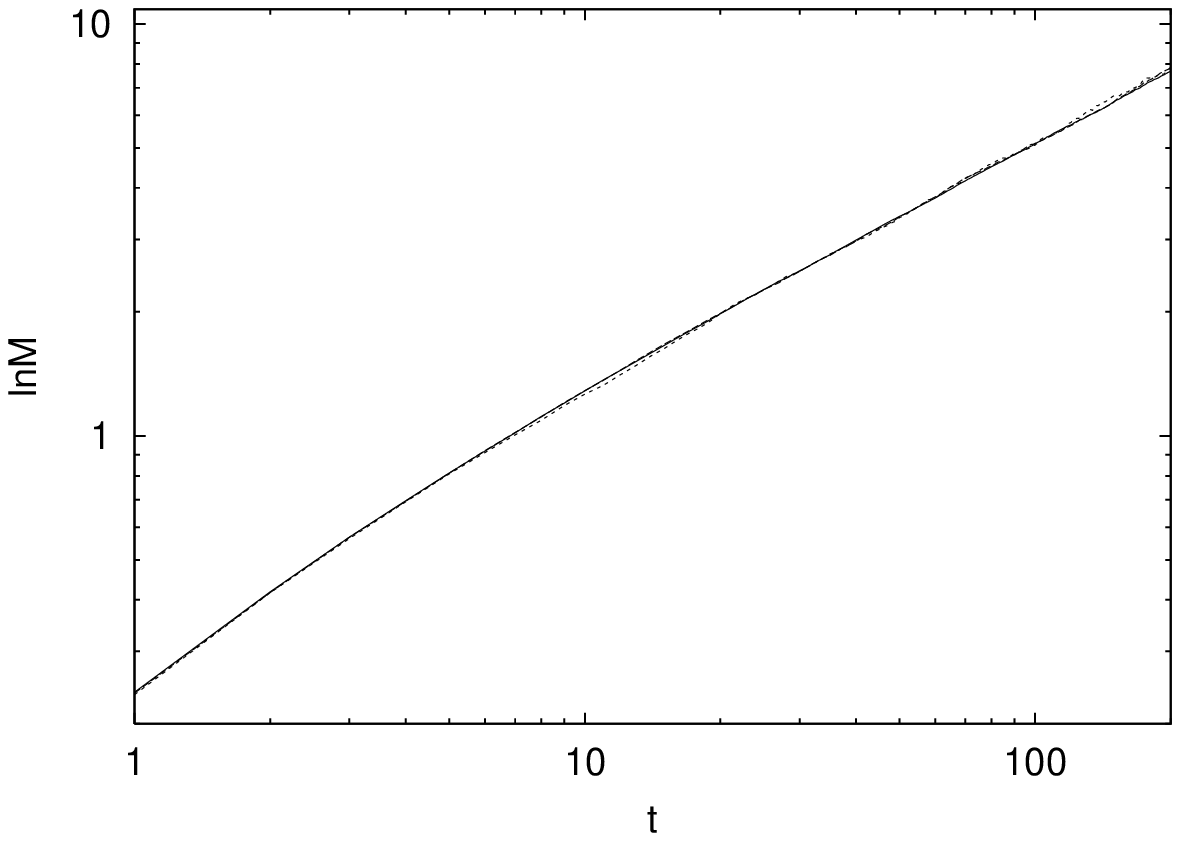}
\end{figure}

\newpage
\begin{figure}[th]
\setlength\epsfxsize{120mm}
\caption{Curves of the second moment of the magnetization
$M^{(2)}(t,L)$ for the Ising model with $m_0 =0$. The solid lines 
present $M^{(2)}(t,L)$ for different lattices of $L$=16,32,64,128
(from top to bottom)
The dotted lines present the rescaled $M^{(2)}(t',L')$. ($\Diamond$
denote rescaling 32 $\rightarrow$ 16, + denote 64 $\rightarrow$ 32 and 
$\Box$ denote 128 $\rightarrow$ 64)}
\epsfbox{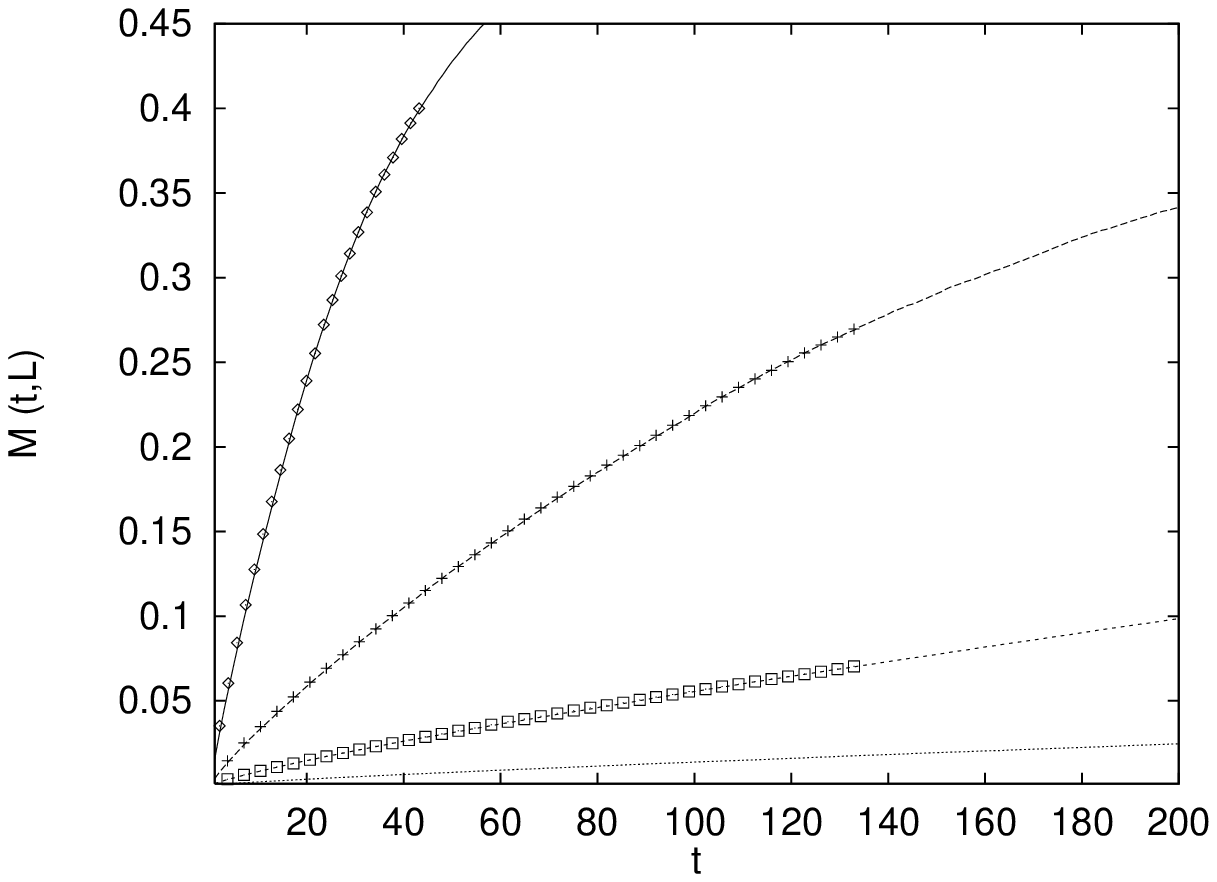}
\vskip 0.4cm
\setlength\epsfxsize{120mm}
\caption{The same as Fig.9, but for the $q$=3 Potts model.~~~~~~~~~~~~~~~~~~~}
\epsfbox{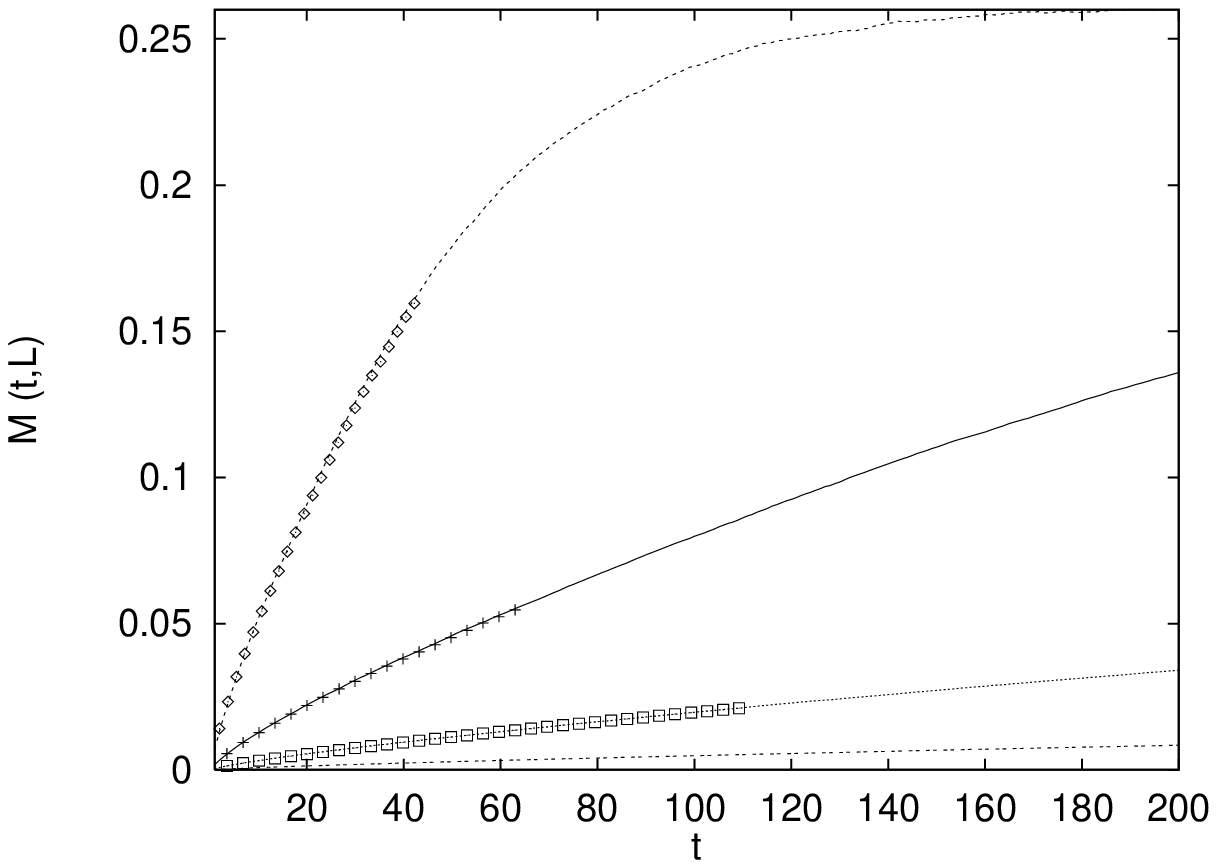}
\end{figure}
 

\baselineskip=12.0pt
\begin{thebibliography}{99}
\bibitem{Bind92}
K. Binder and D.W. Heermann, {\it Monte Carlo Simulation in 
Statistical Physics} (Springer, Berlin, 1992).
\bibitem{Baxter82}
R.J. Baxter, {\it Exactly solved models in statistical mechanics},
(Academic Press, New York, 1982).
\bibitem{Swen87}
R.H. Swendsen and J.S. Wang, Phys. Rev. Lett. {\bf 58}, 86(1987);\\
U. Wolff, Phys. Rev. Lett. {\bf 62}, 361(1989).
\bibitem{Zheng98}
B. Zheng, Int. J. Mod. Phys. {\bf B12}, 1419(1998).
\bibitem{Janss89}
H.K. Janssen, B. Schaub, and B. Schmittmann, Z. Phys. {\bf B73}, 539(1989).
\bibitem{Huse89}
D.A. Huse, Phys. Rev. {\bf B40}, 304(1989).
\bibitem{Li95}
Z.B. Li,  L. Sch\"ulke, and B. Zheng, Phys. Rev. Lett. 
{\bf 74}, 3396(1995); \\Phys. Rev. {\bf E53}, 2940(1996).
\bibitem{Okano97}
K. Okano, L. Sch\"ulke, K. Yamagishi and B. Zheng, Nucl. Phys. 
{\bf B485 [FS]}, 727(1997).
\bibitem{Zheng99}
B. Zheng, M. Schulz and S. Trimper, Phys. Rev. Lett. {\bf 82}, 1891(1999).
\bibitem{Ying98}
H.P. Ying, H.J. Luo, L. Sch\"ulke, and B. Zheng,  
Mod. Phys. Lett. {\bf B12}, 1237(1998). 
\bibitem{Ying00}
H.P. Ying and Kenji Harada, Phys. Rev. {\bf E62}, 174(2000). 
\bibitem{Luo98}
H.J. Luo, L. Sch\"ulke and B. Zheng, Phys. Rev. Lett. 
{\bf 81}, 180(1998).
\bibitem{Okano98}
K. Okano, L. Sch\"ulke, and B. Zheng, Phys. Rev. {\bf D57}, 1411(1998).
\bibitem{Diehl96}
U. Ritschel and H.W. Diehl, Nucl. Phys. {\bf B464}, 512(1996).
\bibitem{Ritsch95}
U. Ritschel and P. Czerner, Phys. Rev. Lett. {\bf 75}, 3882(1995).
\bibitem{Ritsch97}
U. Ritschel and P. Czerner, Phys. Rev. {\bf E55}, 3958(1997).
\bibitem{Hoh77}
P.C. Hohenberg and B.I. Halperin, Rev. Mod. Phys. {\bf 49}, 435(1977).
\bibitem{Ito93}
N. Ito, Physics A {\bf 192}, 604(1993).
\bibitem{Schue95}
L. Sch\"ulke, B. Zheng, Phys. Lett.{\bf A204}, 295(1995).
\bibitem{Luo98a}
H.J. Luo, L. Sch\"ulke and B. Zheng, Phys. Rev. {\bf E57}, 1327(1998).
\bibitem{Zhang99}
J.B. Zhang, L. Wang, D.W. Gu, H.P. Ying and D.R. Ji, 
Phys. Lett. {\bf A262}, 226(1999).
\end{thebibliography}
\end{document}